\documentclass[aps,prb,showpacs,preprint,amssymb,amsmath,floatfix]{revtex4-1}
\pdfoutput=1
\usepackage{graphicx}
\begin{document}
\title{Embedding on to a one-dimensional crystal}
\author{J.E. Inglesfield}
\affiliation{School of Physics and Astronomy,Cardiff University, 
The Parade, Cardiff, CF24 3AA, United Kingdom}
\date{\today}
\begin{abstract}
A simple expression is derived for the band structure of a
one-dimensional periodic potential in terms of two solutions of the
Schr\"odinger equation within the unit cell, one with a
zero-derivative boundary condition on the left-hand end of the cell
and the other with zero derivative on the right-hand end. From this
starting point, a new expression is derived for the embedding
potential -- this can be added to the Hamiltonian for the surface
region of a crystal to replace the semi-infinite substrate, in a
one-dimensional approximation. The results are demonstrated in
calculations of the band structure and embedding potential for Al in
the $[001]$ direction, and the surface electronic structure of the 
Al(001) surface.
\end{abstract}
\pacs{71.15.-m, 71.20.-b, 73.20.-r}
\maketitle
\section{Introduction}
Embedding provides a way of including the effects of the substrate in
a calculation of electronic structure over a restricted region of
space.\cite{jei1} For example, embedding allows us to find the
electronic structure at a surface by solving the Schr\"odinger
equation in the surface region, adding embedding potentials to the
Hamiltonian to include the effects of the semi-infinite substrate and
vacuum regions.\cite{jei2,ishida1} The embedding potentials ensure
that the wave-functions (or the Green function) in the surface region
have the correct boundary corrections on the boundaries of the region,
without matching wave-functions explicitly. The alternative to
embedding is to solve the Schr\"odinger equation for a slab of
material, or a periodic array of slabs. However, slab calculations do
not give the energy continuum of bulk states, and unless the slab is
very thick, the localised surface states interact across the
slab. Embedding calculations have neither of these drawbacks. Several
methods have been developed over the years for finding the embedding
potential for a semi-infinite substrate, with the full crystal
potential, for solving the energy-dependent Schr\"odinger equation in
full-potential calculations of surface electronic
structure.\cite{ishida1,crampin} In this paper we develop a very fast
method for calculating the embedding potential to replace a
semi-infinite one-dimensional periodic potential.

\begin{figure}[h]
\begin{center}
\includegraphics[width=12cm] {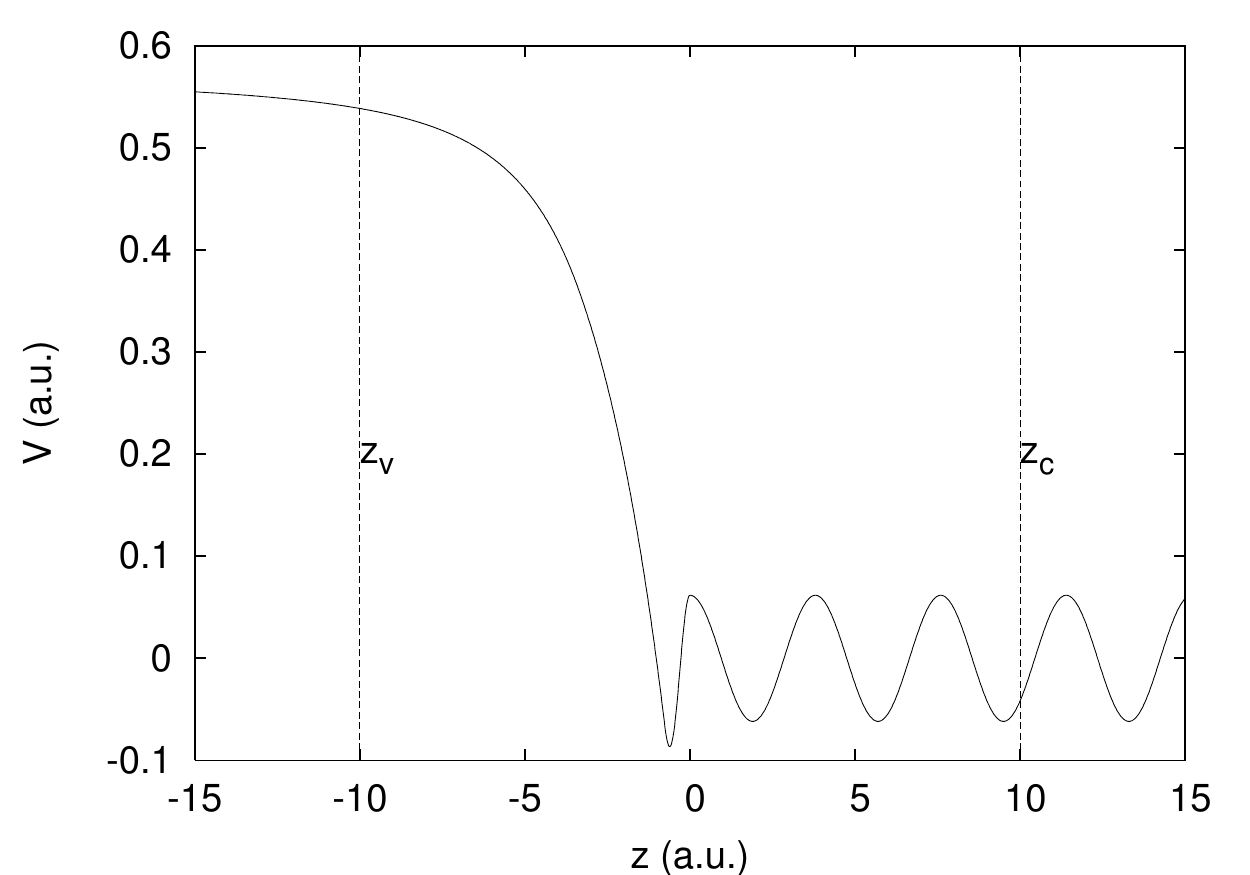}
\end{center}
\caption{One-dimensional potential for modelling the Al(001)
surface.\cite{chulkov1} At $z_v$ the surface region is embedded on to
the vacuum Coulomb tail, and at $z_c$ it is embedded on to the
semi-infinite crystal potential.}
\label{fig1}
\end{figure}
Going from the full three-dimensional crystal potential to a
one-dimensional potential may seem a retrograde step, but Chulkov
\emph{et al.} \cite{chulkov,chulkov1} have recently developed
one-dimensional models of the bulk and surface potential (figure 1),
which give an accurate description of electronic states at the
surface. This has proved particularly useful in many-body and lifetime
studies of Shockley and image potential-induced surface
states.\cite{sarria,ssreport,echenique} This one-dimensional potential
is of course felt by electrons moving in three dimensions, with a
wave-function of the form
\begin{equation}
\Psi(\mathbf{r})=\exp(i\mathbf{K}.\mathbf{R})\psi(z), \label{eq1}
\end{equation}
where $\mathbf{K}$ is the free-electron wave-vector parallel to the
surface, which lies in the $\mathbf{R}$ plane.  Our own interest lies
in using these model surface potentials in the time-dependent
Schr\"odinger equation, which we solve by a new embedding
technique.\cite{jei3} The crystal substrate is replaced by a
time-dependent embedding potential, essentially the Fourier transform
of the energy-dependent embedding potential. This must be evaluated at
a very fine energy grid over a very wide energy range -- hence we need
a method for finding the energy-dependent embedding potential as
efficiently as possible.

The general idea of embedding is that we partition space into two (or
more) regions, solving the Schr\"odinger equation explicitly in what
we call region I, with the rest of space, region II, replaced by an
embedding potential added to the Hamiltonian of region I. The
embedding potential is defined over the interface $S$ between regions
I and II.\cite{jei1} Using a variational method, it can be shown that
the wave-function for the system, $\psi(\mathbf{r})$, satisfies the
following Schr\"odinger equation in region I,
\begin{eqnarray}
\lefteqn{\left(-\frac{1}{2}\nabla^2+V(\mathbf r)\right)
\psi(\mathbf{r})+\delta(\mathbf{r}-\mathbf{r}_S)\left[\frac{1}{2}
\frac{\partial\psi}{\partial n_S}\right.}\nonumber\\&&+
\left.\int_S d\mathbf{r}_S'\left(\Sigma(\mathbf{r}_S,\mathbf{r}_S';
\epsilon)+(E-\epsilon)\frac{\partial \Sigma}{\partial \epsilon}
\right)\psi(\mathbf{r}'_S)\right]=E\psi(\mathbf{r}),\label{eq2}
\end{eqnarray}
where the integral is over $S$.\cite{jei1} The embedding potential
$\Sigma$ is a function of two interface coordinates and is evaluated
at trial energy $\epsilon$; the energy derivative term gives $\Sigma$
at the required energy $E$, to first order in $(E-\epsilon)$. The
embedding potential gives the generalised logarithmic derivative over
$S$ of the wave-function in region II,\cite{jei2}
\begin{equation}
\frac{\partial\psi(\mathbf{r}_S)}{\partial n_S}=-2\int_S
d\mathbf{r}'_S \Sigma(\mathbf{r}_S,\mathbf{r}'_S;E)
\psi(\mathbf{r}'_S),\label{eq3}
\end{equation}
and this ensures that $\psi$ in region I matches correctly in
amplitude and derivative on to the solution in II. In
three-dimensional applications, we find $\Sigma$ from the Green
function for region II satisfying the zero-derivative boundary
condition on $S$.\cite{jei1,jei2} However, in the one-dimensional
application described in this paper, we use Eq.~\ref{eq3} to find the
embedding potential -- in one-dimension, this becomes a relation
between amplitude and derivative at the point between regions I and II.

As the use of $\Sigma$ suggests, the embedding potential is a form of
self-energy, an energy-dependent, possibly complex potential added on
to the Hamiltonian to replace a region of phase space or Hilbert
space. In fact the embedding potential in the linear combination of
atomic orbitals formalism \cite{baraff} is usually called the
self-energy, often in connection with conduction through linear
molecules attached at each end to metallic
contacts.\cite{mujica,starikov} The contacts and associated electron
reservoirs are replaced in the tight-binding Hamiltonian of the system
by self-energies (embedding potentials in tight-binding
guise).\cite{davies,ishida2} In our embedding method, in which space
is partitioned, we can use any convenient basis set for expanding the
wave-function or Green function in region I.

We now briefly describe the structure of this paper. In section II we
shall use solutions of the Schr\"odinger equation in one unit cell to
calculate the complex band structure, which describes the allowed
solutions for the semi-infinite crystal at energy $E$, and the
logarithmic derivative of these solutions, hence the embedding
potential $\Sigma(E)$. Related results have been derived by
Butti,\cite{butti} using a different method, and his expression for
the one-dimensional embedding potential was used in a recent analysis
of electron spectroscopies from adsorbates.\cite{trioni} The result
for the band structure has previously been derived by Kohn,\cite{kohn}
and with particular reference to the sinusoidal potential (the Mathieu
problem), by McLachlan.\cite{mclachlan} The expressions are remarkably
simple, and the band structure formula in particular may be useful in
teaching, where one-dimensional potentials such as the Kronig-Penney
model frequently serve as an introduction to band
theory.\cite{merzbacher} In section III we shall illustrate the use of
the embedding potential in a calculation of the density of states at
the Al(001) surface using the one-dimensional model potential of
Chulkov \emph{et al}.\cite{chulkov1} The expression for the embedding
potential of a one-dimensional crystal is useful not only in the
energy domain, but we are also using it in Fourier transform (with an
extra factor of $1/E$)\cite{jei3} to study time-dependent processes at
surfaces, such as electron emission.

Atomic units are used in this paper, with $e^2=\hbar=m_e=1$.
\section{Band structure and embedding potential}
Our starting point for finding the band structure and embedding
potential is the Green function formula for the wave-function
$\psi(z)$ in some interval in terms of the derivatives at the ends of
the range,
\begin{equation}
\psi(z)=\frac{1}{2}[G(z,a)\psi'(a)-G(z,0)\psi'(0)],
\;\;0\le z\le a,\label{eq5}
\end{equation}
where we take the range to be the unit cell between $z=0$ and $z=a$ of
the infinite one-dimensional crystal; $G$ is the Green function with
zero-derivative boundary conditions at the ends of the unit cell. This
formula is analogous to the equation in electrostatics giving the
potential inside some region of space in terms of the boundary values
of the electric field,\cite{jackson} and can be derived in exactly
the same way.

This equation satisfied by the wave-function within the unit cell does
not depend on $\psi(z)$ in the rest of the crystal, but we now impose
the Bloch form of wave-function. Evaluating $\psi$ at $z=0$,
Eq.~\ref{eq5} becomes
\begin{equation}
\psi(0)=\frac{1}{2}\psi'(0)[G(0,a)\exp(ika)-G(0,0)],
\label{eq9}
\end{equation}
where $k$ is the Bloch wave-vector. But we can also evaluate
Eq.~\ref{eq5} at $z=a$, giving
\begin{equation}
\psi(a)=\frac{1}{2}\psi'(a)[G(a,a)-G(a,0)\exp(-ika)].\label{eq10}
\end{equation}
Now the logarithmic derivative $\psi'/\psi$ is invariant to a lattice
displacement, so comparing these two equations we obtain
\begin{equation}
G(0,a)\exp(ika)-G(0,0)=G(a,a)-G(a,0)\exp(-ika).\label{eq11}
\end{equation}
The Green function is symmetric in its spatial variables, so
$G(0,a)=G(a,0)$, and Eq.~\ref{eq11} simplifies to
\begin{equation}
\cos(ka)=\frac{G(0,0)+G(a,a)}{2G(0,a)}.\label{eq12}
\end{equation}  
This can be simplified further by using the expression for the Green
function in terms of wave-functions $\phi_1(z)$ and $\phi_2(z)$, which
satisfy the boundary conditions of the Green function at each end of
the unit cell: $\phi_1$ satisfies the Schr\"odinger equation with
$\phi_1'(0)=0$, and at the other end of the unit cell $\phi_2'(a)=0$,
\begin{equation}
G(z,z')=-\frac{2\phi_1(z_<)\phi_2(z_>)}{W(\phi_1,\phi_2)},\label{eq13}
\end{equation}
where $W$ is the Wronskian.\cite{morse} Furthermore, we take
$\phi_1(0)=1$, $\phi_2(a)=1$. Then substituting Eq.~\ref{eq13} into
Eq.~\ref{eq12} gives the remarkably simple expression for the
wave-vector,
\begin{equation}
\cos(ka)=\frac{\phi_1(a)+\phi_2(0)}{2}.\label{eq14}
\end{equation}
So we can find the Bloch wave-vector corresponding to a particular
energy by two integrations through the unit cell, one in each
direction, to find $\phi_1$ and $\phi_2$. If the potential is
symmetric with respect to the origin, $\phi_1(a)=\phi_2(0)$, and the
result simplifies to
\begin{equation}
\cos(ka)=\phi_1(a), \label{abram} 
\end{equation}

Using wave-function matching, Kohn \cite{kohn} gives a result related to
Eq.~\ref{eq14}, though instead of using $\phi_2$, his second
wave-function has zero amplitude at $z=0$ and unit
derivative. Abramowitz and Stegun \cite{abramowitz} quote
Eq.~\ref{abram} in section 20.3.10 in the chapter on Mathieu
functions, and this is proved by McLachlan \cite{mclachlan} using a
Fourier expansion of the wave-function. An alternative approach is
given by Butti:\cite{butti} he uses the same starting
wave-functions as Kohn to find the transfer matrix; this relates
solutions of the Schr\"odinger equation at each end of the unit cell,
and its eigenvalues give the Bloch phase factors.

If the solutions $\phi_1$ and $\phi_2$ at real energy $E$ satisfy
$|\phi_1(a)+\phi_2(0)|/2> 1$, then $E$ lies in a bulk band-gap, and
the wave-vector $k$ satisfying Eq.~\ref{eq14} is complex (either pure
imaginary, or complex with a real part at the Brillouin zone
boundary). This corresponds to a Bloch solution forbidden in the
infinite bulk crystal, but allowed in the semi-infinite crystal, for
example in the case of a crystal with a surface.\cite{fauster} More
generally we can find the band-structure at \emph{complex} $E$ -- we
shall need this in section III. In this case, $k$ satisfying
Eq.~\ref{eq14} is always complex, and the two solutions correspond to
two different physical cases. One solution corresponds to waves
travelling to the right and decaying in this direction, with
$\mbox{Im}\,k>0$; the other solution, with $\mbox{Im}\,k<0$ is
travelling and decaying to the left.  Which solution we take depends
on the particular problem -- in our case, with the semi-infinite bulk
crystal lying in the positive $z$-direction (figure \ref{fig1}) the
physical solution corresponds to $\mbox{Im}\,k>0$.

\begin{figure}[h]
\begin{center}
\includegraphics [width=12cm] {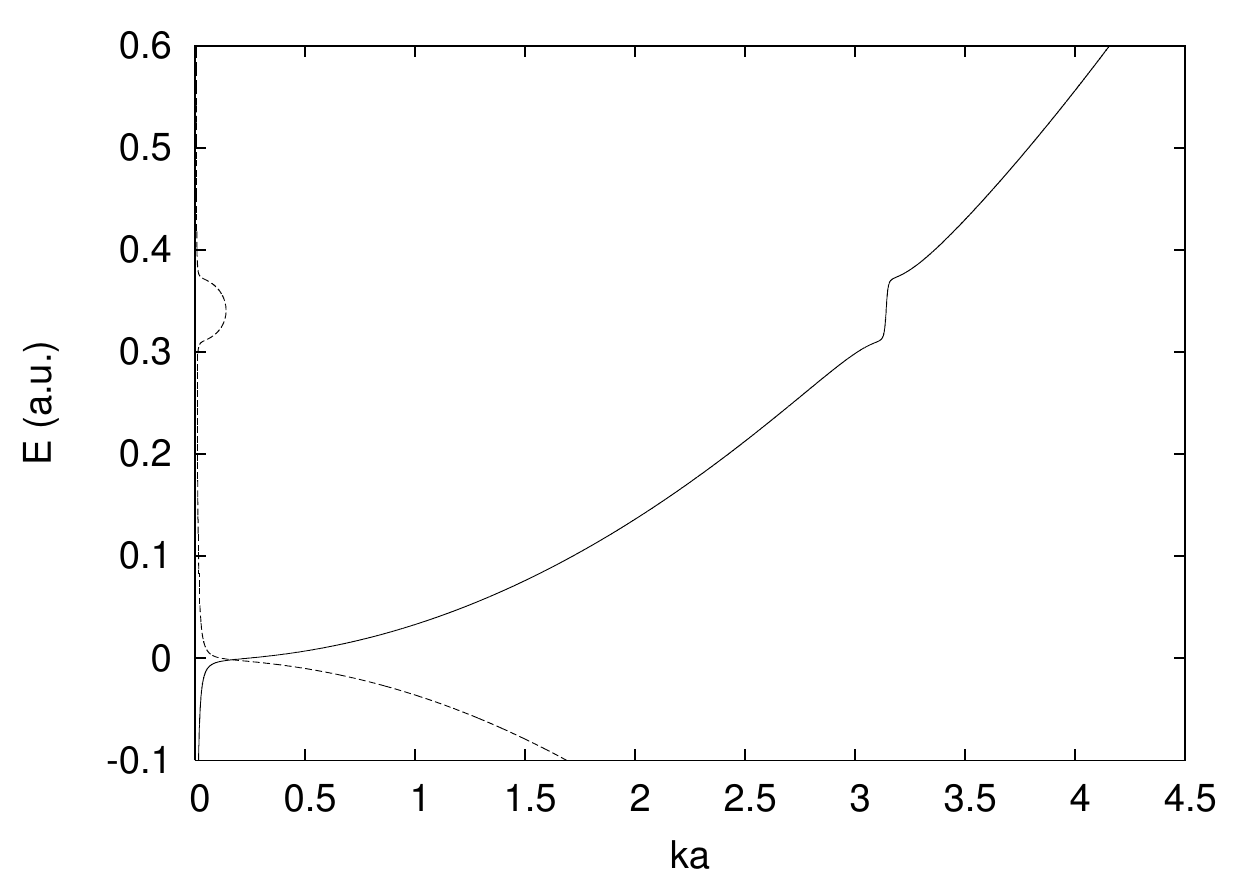}
\end{center}
\caption{Complex band structure for Al in the $[001]$ direction 
with the one-dimensional potential. $\mbox{Re}\,E$ is plotted along
the $y$-axis, and $\mbox{Im}\,E$ is kept fixed at $0.002$ a.u. Solid
line, $\mbox{Re}\,k$; dashed line, $\mbox{Im}\,k$.}
\label{fig2}
\end{figure}

To illustrate the method, we evaluate the complex band structure of a
one-dimensional pseudopotential corresponding to Al in the $[001]$
direction,
\begin{equation}
V(z)=A\cos(2\pi z/a),\label{eq19}
\end{equation}
where the amplitude is given by $A=0.0618$ a.u. and the lattice
constant $a$ is $3.8$ a.u. -- the bulk potential of figure
\ref{fig1}.\cite{chulkov1} This is the Mathieu problem, but our method
holds for arbitrary potential and arbitrary origin. We calculate the
band structure at complex energy, keeping the imaginary part of the
energy fixed at 0.002 a.u. At each energy we find $\phi_1$ and
$\phi_2$ by integrating the Schr\"odinger equation through the unit
cell using Numerov's method,\cite{jos} with a spatial interval of
0.002 a.u., starting off the integrations using the method described
by Quiroz Gonz\'{a}les and Thompson,\cite{quiroz} The resulting band
structure, smoothed by working with the small imaginary energy, is shown
in figure \ref{fig2}. $\mbox{Re}\,k$ is plotted in the extended zone
scheme to give a continuous curve, weaving its way from band to band
around the bulk band gaps; as we expect, $\mbox{Im}\,k$ is non-zero
below the bottom of the band, and in the band gaps. Our analysis of
the complex band structure is important now that we turn to the
embedding potential.

We need the embedding potential $\Sigma_c$ to replace the
semi-infinite bulk crystal, which we assume lies to the right of $z_c$
(figure \ref{fig1}), where we now put the origin of the unit cell. In
the one-dimensional case, Eq.~\ref{eq3} simplifies to the logarithmic
derivative,
\begin{equation} 
\Sigma_c=-\frac{1}{2}\frac{\psi'(0)}{\psi(0)}, \label{eq20}
\end{equation}
where $\psi$ is the solution of the Schr\"odinger equation for the
semi-infinite crystal travelling or decaying as
$z\rightarrow\infty$.\cite{jei4} We can find this logarithmic
derivative directly from Eq.~\ref{eq9}, and substituting the Wronskian
expression for the Green function, Eq.~\ref{eq13}, we obtain the
following result for the embedding potential,
\begin{equation}
\Sigma_c=\frac{W(\phi_1,\phi_2)}{2[\exp(ika)-\phi_2(0)]}.
\label{eq21}
\end{equation}
The wave-vector $k$ in this expression corresponds to the wave
travelling or decaying to the right -- this is precisely the value of
$k$ which was discussed above in the band structure calculation
(figure \ref{fig2}). This result is different in form from the
expression derived by Butti,\cite{butti} but the two must be
identical.

\begin{figure}[h]
\begin{center}
\includegraphics [width=12cm] {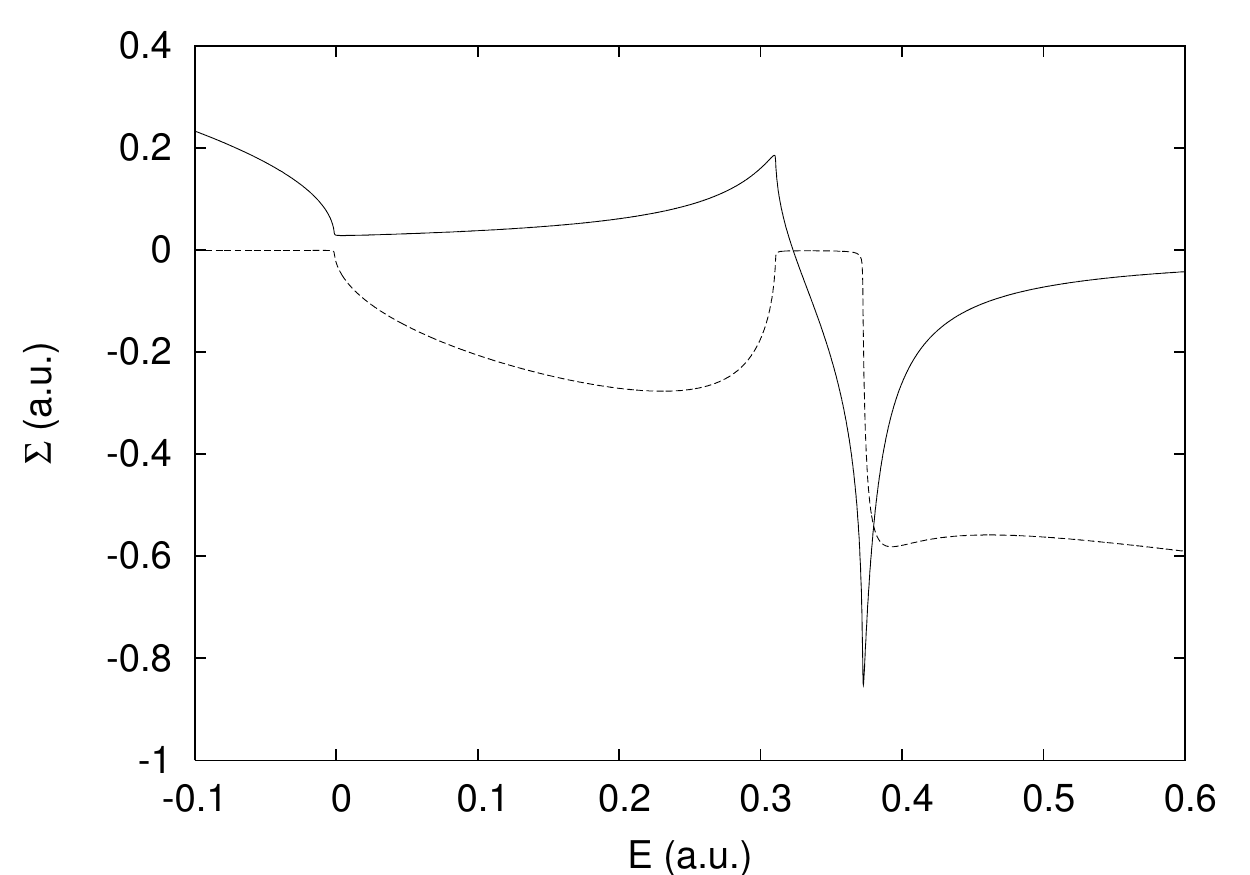}
\end{center}
\caption{Embedding potential for a bulk Al substrate with the
one-dimensional potential, evaluated at $z_c=10$ a.u., as a function
of $\mbox{Re}\,E$. $\mbox{Im}\,E$ is kept fixed at $0.0002$ a.u. Solid
line, $\mbox{Re}\,\Sigma_c$; dashed line, $\mbox{Im}\,\Sigma_c$.}
\label{fig3}
\end{figure}
We now calculate the embedding potential for the Al substrate at
$z_c=10$ a.u., at a complex energy with an imaginary part equal to
0.0002 a.u. The only additional feature compared with the band
structure calculation is that the expression for the embedding
potential involves the Wronskian, in which we evaluate the
derivatives by finite differences. The results are shown in figure
\ref{fig3}, and we note that $\mbox{Im}\,\Sigma$ is negative, as we
require from causality.

\section{Surface density of states}
The Al(001) surface is suitable for demonstrating this embedding
potential -- the one-dimensional crystal potential of Chulkov 
\emph{et al.} \cite{chulkov1} should work well in this s-p bonded
metal, and the (001) surface shows a Shockley surface
state,\cite{greg} as well as structure induced by the image
potential.\cite{maziar2}

We shall calculate the local density of states $\sigma(z,\varepsilon)$
in the surface region, the charge density of states with energy
$\varepsilon$. This is given by the sum over states
\begin{equation}
\sigma(z,\varepsilon)=\sum_i|\psi_i(z)|^2\delta(\varepsilon-
\varepsilon_i),\label{eq23}
\end{equation}
where $\psi_i(z)$ is a wave-function of the system with energy
$\varepsilon_i$, and it can be written in terms of the Green function
evaluated at an energy with a small imaginary part,
\begin{equation}
\sigma(z,\varepsilon)=\frac{1}{\pi}\mbox{Im}\,G(z,z;\varepsilon+i\eta).
\label{eq24}
\end{equation}
(Note that we use $\varepsilon$ to denote a real energy, and $E$ to
denote an energy which may be complex -- here, $E=\varepsilon+i\eta$.)

Let us take the surface region, where we calculate $G$, between $z_v$
on the vacuum side and $z_c$ on the crystal side (figure \ref{fig1}),
and then in this region $G$ satisfies the Schr\"odinger equation
embedded on both sides,
\begin{eqnarray}
\lefteqn{-\frac{1}{2}\frac{\partial^2 G}{\partial z^2}+
(V-E)G(z,z';E)+\delta(z-z_c)\left[\frac{1}{2}\frac{\partial G}
{\partial z}+\Sigma_c(E)G(z_c,z';E)\right]}\hspace*{3cm}\nonumber\\&&
\hspace*{-2cm}+\delta(z-z_v)\left[-\frac{1}{2}\frac{\partial G}
{\partial z}+\Sigma_v(E)G(z_v,z';E)\right]=\delta(z-z').\label{eq25}
\end{eqnarray}
The surface region is embedded at $z_c$ on to the crystal, with the
crystal embedding potential $\Sigma_c$, and at $z_v$ on to the vacuum
region, which is replaced by $\Sigma_v$. Unlike Eq.~\ref{eq2},
the embedded Schr\"odinger equation for the Green function does not
contain the energy derivative of the embedding potentials; this is
because we know the energy at which they should be evaluated -- the
energy $E$ of the Green function.

\begin{figure}[h]
\begin{center}
\includegraphics [width=14cm] {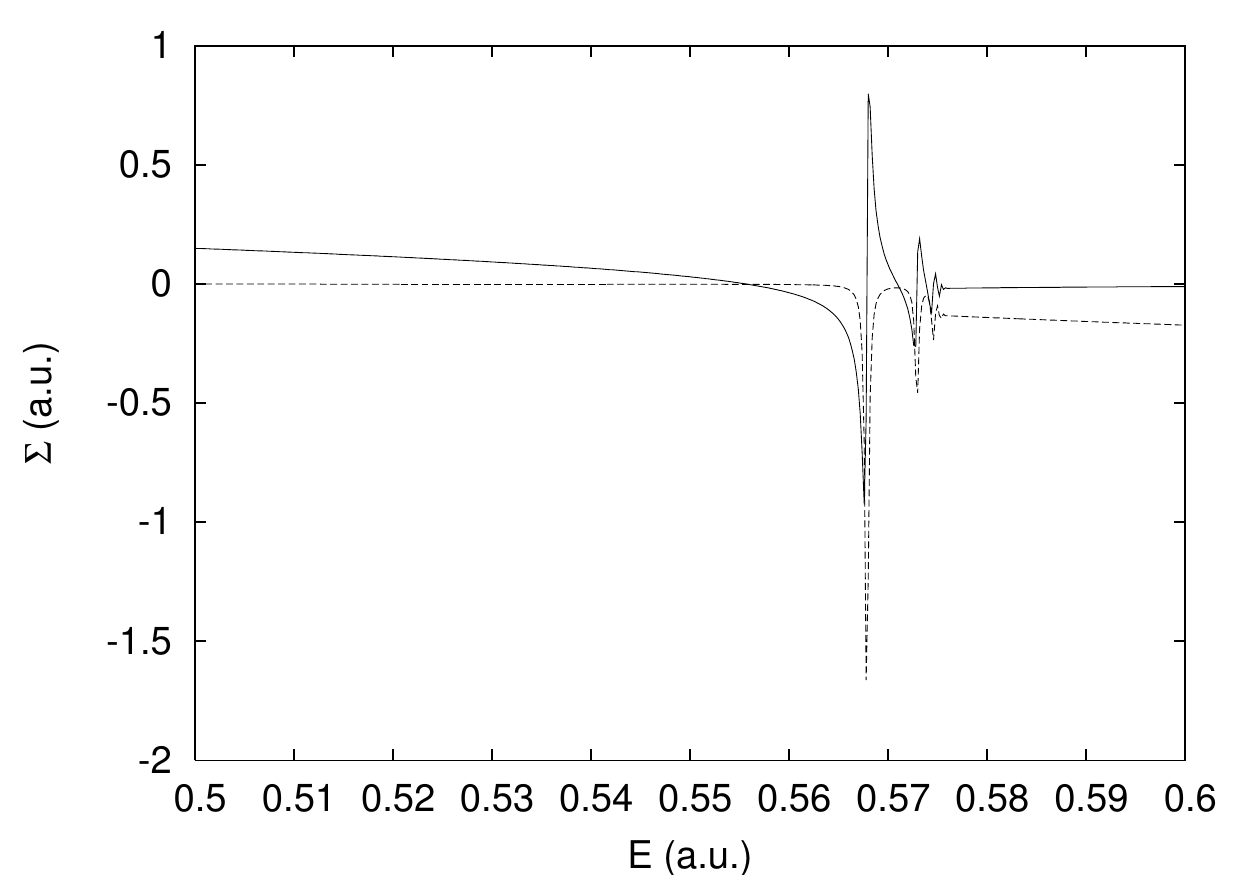}
\end{center}
\caption{Embedding potential for the vacuum region, evaluated at 
$z_v=-10$ a.u. with the image plane at $z_0=-3.44$ a.u., as a function
of $\mbox{Re}\,E$. $\mbox{Im}\,E$ is kept fixed at $0.0002$ a.u. Solid
line, $\mbox{Re}\,\Sigma_v$; dashed line, $\mbox{Im}\,\Sigma_v$.}
\label{fig4}
\end{figure}
Although it is not the main topic of this paper, we shall say a few
words about the embedding potential $\Sigma_v$, which replaces the
vacuum region outside the crystal. Here the electron feels the Coulomb
tail of the image potential,
\begin{equation}
V(z)=V_0-\frac{1}{4|z_0-z|},\label{eq26}
\end{equation}
where $V_0$ is the vacuum zero (we take the average potential in the
bulk crystal as the zero of energy), and $z_0$ is the position of the
image plane. Again we calculate the embedding potential from the
logarithmic derivative of the outgoing or decaying solution of the
Schr\"odinger equation -- this is a combination of the regular and
irregular Coulomb functions\cite{maziar1,thompson} $F_0$ and $G_0$ (in the notation of Abramowitz and Stegun\cite{abramowitz}), with angular momentum $L=0$

\begin{equation}
\psi(z)=H_0^-(\eta,\rho)=G_0(\eta,\rho)-iF_0(\eta,\rho) \label{eq28}
\end{equation}
with arguments given by
\begin{eqnarray}
\rho&=&\sqrt{2(E-V_0)}(z-z_0) \nonumber\\ 
\eta&=&\frac{1}{4\sqrt{2(E-V_0)}}. \label{eq27}
\end{eqnarray}
Thompson and Barnett \cite{thompson} give a rapidly converging
continued fraction expression for $H_0^{-\prime}/H_0^-$, from which we
can immediately find the vacuum region embedding potential
$\Sigma_v$. Figure \ref{fig4} shows $\Sigma_v$, at $z_v=-10$ a.u. with
the image plane at $z_0=-3.44$ a.u. as a function of energy, taking
$\mbox{Im}\,E=0.0002$ a.u. The structure just below the vacuum zero at
$E=0.577$ a.u. comes from bound states of the Coulomb potential.

We are now in a position to solve the embedded Schr\"odinger equation
in region I, the near-surface region, using a basis set expansion for
$G$,
\begin{equation}
G(z,z';E)=\sum_{i,j}G_{ij}(E)\chi_i(z)\chi_j(z'). \label{eq29}
\end{equation}
The Schr\"odinger equation Eq.~\ref{eq25} then reduces to a matrix
equation,
\begin{equation}
\sum_j\left(H_{ij}+\Sigma_{ij}(E)-ES_{ij}\right)G_{jk}=\delta_{ik},
\label{eq30}
\end{equation}
where the Hamiltonian matrix is given by
\begin{equation}
H_{ij}=\frac{1}{2}\int_{z_v}^{z_c}dz\frac{d\chi_i}{dz}\frac{d\chi_j}
{dz}+\int_{z_v}^{z_c}dz\chi_i(z)V(z)\chi)_j(z), \label{eq31}
\end{equation}
the embedding matrix by
\begin{equation}
\Sigma_{ij}(E)=\Sigma_c(E)\chi_i(z_c)\chi_j(z_c)+\Sigma_v(E)\chi_i(z_v)
\chi_j(z_v), \label{eq32}
\end{equation}
and $S$ is the overlap matrix,
\begin{equation}
S_{ij}=\int_{z_v}^{z_c}dz\chi_i(z)\chi_j(z). \label{eq33}
\end{equation}
We usually use trigonometric basis functions,
\begin{equation}
\chi_m(z)=\left\{\begin{array}{l}\cos\frac{m\pi\zeta}{2D},\,
\;m\mbox{ even}\\\sin\frac{m\pi\zeta}{2D},\;\;m\mbox{ odd}\end{array}
\right., \label{eq34}
\end{equation}
where $\zeta$ is measured from the mid-point of region I,
\begin{equation}
\zeta=z-\frac{z_v+z_c}{2}\label{eq35}
\end{equation}
and a value of $D>(z_c-z_v)/2$ gives a range of logarithmic derivatives at
$z_v$ and $z_c$ for matching on to the embedding potentials. Solving
Eq.~\ref{eq30} then gives us $G$, and via Eq.~\ref{eq23} the local
density of states.

\begin{figure}[h]
\begin{center}
\includegraphics [width=14cm] {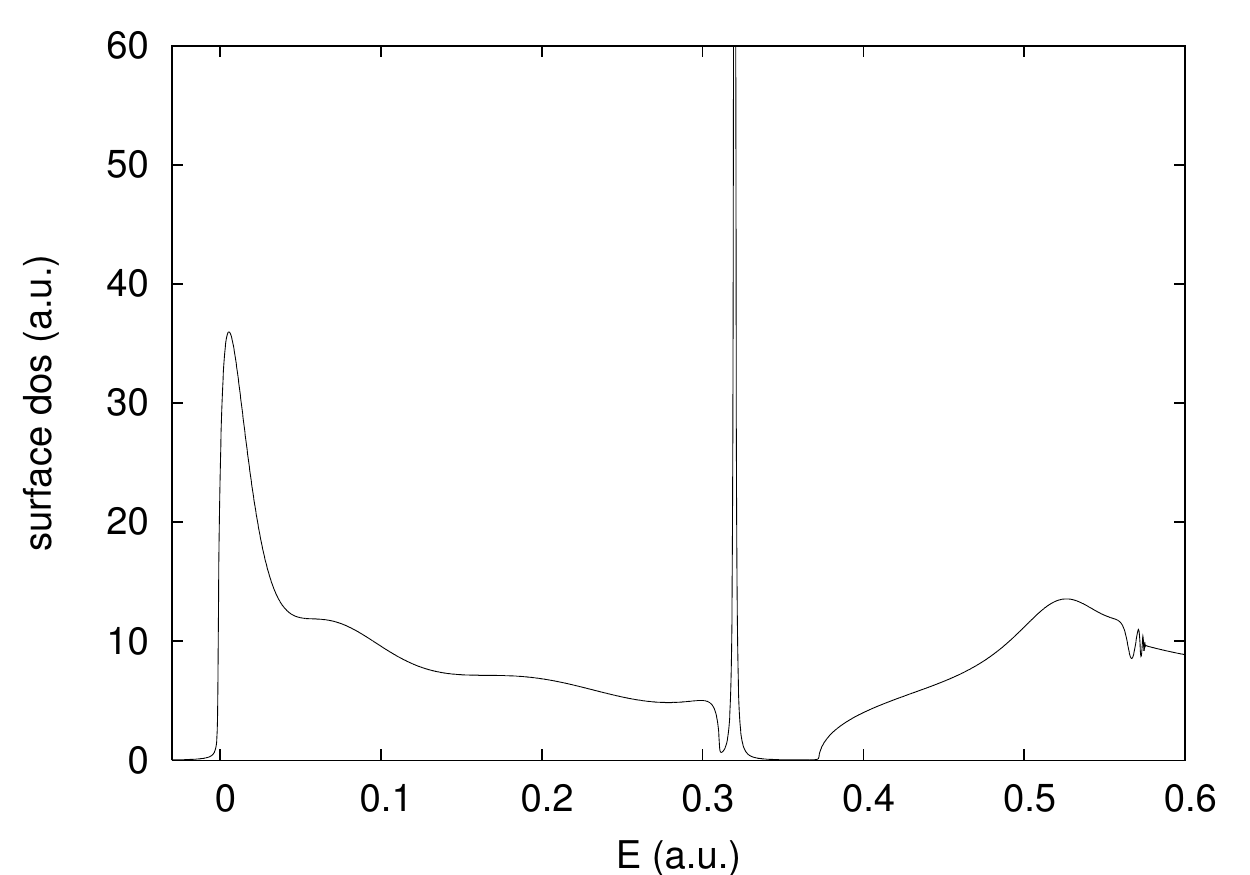}
\end{center}
\caption{Surface density of states of Al(001), the local density of
states integrated through region I, taken between $z_v=-10$ a.u. and
$z_c=+10$ a.u.}
\label{fig5}
\end{figure}
Integrating the local density of states over the surface region, we
obtain the surface density of states for Al(001), shown in figure
\ref{fig5}. Here we take region I between $z_v=-10$ a.u. and $z_c=+10$
a.u., and 20 basis functions are used, defined with $D=13$ a.u. These
results, which correspond to $\mathbf{K}=0$ in three dimensions, show
the Shockley surface state near the bottom of the band gap, broadened
by the imaginary part of the energy at which the the Green function
and embedding potentials are evaluated, here taken to be $0.0002$
a.u. Less familiar is the structure just below the vacuum edge, at
$E=0.577$ a.u., which comes from surface resonances induced by the
image potential.\cite{maziar2} These are the image potential surface
states, broadened into resonances by interacting with the continuum of
bulk states. All the features in the surface electronic structure --
the continuum of bulk states, the Shockley and image surface states --
depend on a correct treatment of the substrate and vacuum regions, via
the embedding potentials.
\section{Concluding remarks}
The motivation for this work was to derive the embedding potential for
a one-dimensional model of a crystal as efficiently as possible, for
subsequent use in energy-dependent or time-dependent surface
calculations. We have shown how the band structure and embedding
potential can be found by integrating through the unit cell at a
chosen energy, starting at each end of the cell with a zero-derivative
boundary condition. Using these results, we have calculated the band
structure and embedding potential for Al in the [001] direction as a
function of complex energy, and applied this to calculating the
surface density of states at the Al(001) surface.

We are now using these results in studies of time-dependent excitation
processes at surfaces, taking the Fourier transform of the embedding
potential to find the time-dependent embedding potential. But
one-dimensional periodic potentials occur in a variety of contexts --
not least in teaching -- and we hope that the results and their derivations will be of
wider
interest. 
\section*{Acknowledgements}
I am grateful to Greg Benesh (Baylor) for his useful comments
on the manuscript, to Eugene Chulkov (San Sebasti\'{a}n) for help with
his one-dimensional potential, to Simon Crampin (Bath) and Hiroshi
Ishida (Tokyo) for their references to Abramowitz and Stegun and Kohn's
article, to Jos Thijssen (Delft) for suggestions about wave-function
matching, and to Mario Trioni (Milano-Bicocca) who drew my attention to the work of Gabriele Butti.

\end{document}